\documentstyle[aps,prd]{revtex}
\begin{document}
\draft
\title{The Head-On Collision of Two Equal Mass Black Holes: \\
       Numerical Methods}

\author{Peter Anninos${}^{(1)}$, David
Hobill${}^{(1,2)}$, Edward Seidel${}^{(1,4)}$, Larry Smarr${}^{(1,4)}$
and Wai-Mo Suen${}^{(3)}$}

\address{${}^{(1)}$ National Center for Supercomputing Applications \\ 605
E. Springfield Ave., Champaign, IL 61820}

\address{${}^{(2)}$ Department of Physics \\
University of Illinois, Urbana, IL 61801}

\address{${}^{(3)}$ Department of Physics and Astronomy \\ University of
Calgary, Calgary, Alberta, Canada T2N 1N4}

\address{${}^{(4)}$ Department of Physics \\
Washington University, St. Louis, Missouri, 63130}

\date{\today}
\maketitle
\begin{abstract}
The head-on collision of two nonrotating axisymmetric equal mass black
holes is treated numerically. We take as initial data the
single parameter family of time-symmetric solutions
discovered by Misner which consists of two Einstein-Rosen bridges
that can be placed arbitrarily distant from one another.
A number of problems associated with previous attempts to evolve these
data sets have been overcome.  In this article, we
discuss our choices for coordinate
systems, gauges, and the numerical algorithms
that we have developed to evolve this
system.
\end{abstract}



\section{introduction}
\label{sec:introduction}

The coalescence of two black holes is considered to be one of the
most promising sources of gravitational waves~\cite{Thorne94a}.
In a series of papers
{}~\cite{Anninos93b,Anninos94b,Anninos93a}
we investigate numerically a special case of the black hole
coalescence problem, namely the head-on collision of two
equal mass black holes.
Numerical computations are extremely difficult due to
coordinate singularities, large gradients in the metric
components, and numerical instabilities inherent in the two black hole
spacetimes.
Building on the work of Dewitt, Cadez, Smarr, and Eppley
{}~\cite{Cadez71,Smarr75,Eppley75,Smarr79,Smarr76}
(henceforth abbreviated as DCSE)
and the more recent work involving distorted, single black
holes~\cite{Anninos93c,Bernstein93b,Abrahams92a}, many of these
numerical problems have been overcome in the present work.

Our work (like that of DCSE) is based on studying evolutions of the
Misner initial data set~\cite{Misner60}, a single parameter family of
time symmetric solutions that allows the
initial separation between the two black holes and the total
ADM mass of the system to be specified.
Section \ref{sec:initial} discusses this
data set briefly.  In section \ref{sec:theoretical} we present the
basic equations, formalism, coordinate system and gauge
considerations.  Section \ref{sec:coordinate} details some of the
various numerical problems that we encountered along with the
computational methods we developed to overcome them, drawing parallels
between our methods to those of DCSE when appropriate.  Section
\ref{sec:methods} reviews the general numerical methods we use to
integrate the discrete Einstein equations.  We also present
convergence studies testing the reliability and robustness of our
code.

Since this paper is devoted primarily to numerical methods and
tests of our code, we refrain from presenting results such as
extracted waveforms, energy radiated, horizon oscillations, {\it etc.},
except where we discuss convergence results for our code. Instead we
refer the reader to the series of related papers
\cite{Anninos93b,Anninos94b,Anninos93a}, for a detailed analysis
of the physics of colliding black holes.

\section{the Misner initial data}
\label{sec:initial}

There are a number of ways to construct initial data representing two
black holes. One of the simplest to work with is the single parameter
family of analytic data derived by Misner~\cite{Misner60} for the
Einstein-Rosen~\cite{Einstein35} model of two asymptotically flat
sheets joined by two throats.  Detailed studies of the
Einstein-Rosen bridge construction and variations of it were
discussed by Misner~\cite{Misner63}, Lindquist~\cite{Lindquist63},
Brill and Lindquist~\cite{Brill63}, and others.  Data sets of this
type were first investigated numerically by Hahn and
Lindquist~\cite{Hahn64}, and later by DCSE.

The spatial 3-metric for the
Misner data is written in the following conformally
flat form using cylindrical coordinates
\begin{equation}
dl^2=\Psi_M^4\left(d\rho^2+dz^2+\rho^2d\phi^2\right).
\label{mmetriczr}
\end{equation}
The conformal factor
$\Psi_M$ defined by
\begin{equation}
\Psi_M=1+\sum_{n=1}^{\infty}\frac{1}{\sinh(n\mu)}
       \left(\frac{1}{{}^{+}r_n}+\frac{1}{{}^{-}r_n}
       \right),
\label{mPsizr1}
\end{equation}
and
\begin{equation}
{}^{\pm} r_n=\sqrt{\rho^2+\left[z\pm \coth(n\mu)\right]^2}
\label{mPsizr2}
\end{equation}
solves the Hamiltonian constraint (\ref{ham})
with the proper isometry imposed between the upper and lower sheets.
This data set is both
axisymmetric and time symmetric ($K_{ij}=0$) and represents two equal
mass black holes with zero rotation.  The two black hole centers are
aligned along the axis of symmetry ($z$-axis) so the physical
interaction is a head-on collision along the axis.
The free parameter $\mu$ is related to the physical parameters
$M$ (half the total ADM mass
and approximately the mass of a single black hole when the holes
are ``infinitely'' separated)
\begin{equation}
M=M_{ADM}/2=2\sum_{n=1}^{\infty}\frac{1}{\sinh(n\mu)},
\label{tmass}
\end{equation}
and $L$ (the proper
distance along the spacelike geodesic connecting the throats)
\begin{equation}
L=2\left[1+2\mu\sum_{n=1}^{\infty}\frac{n}{\sinh(n\mu)}\right].
\end{equation}
The effect of increasing $\mu$ is to set the two black holes further
away from one another and decrease the total mass of the system.

Comparing with the
isotropic form of the Schwarzschild metric
\begin{equation}
ds^2=-\left(\frac{1-m/2r}{1+m/2r}\right)^2 dt^2
     +\left(1+\frac{m}{2r}\right)^4 \left(dx^2+dy^2+dz^2\right),
\label{smetric}
\end{equation}
we can see how the Misner data (\ref{mmetriczr})
is similar to (\ref{smetric}). If one
associates $m_1 \sim \sum 2/\sinh{n\mu}$ as the mass of a single
black hole when the two holes are separated by large distances ($\mu
\rightarrow
\infty$), then
the Misner data resembles (\ref{smetric}) if we identify $m=m_1$ for
regions close to one of the throats and $m=2m_1$ in the far field.
(See also, for example, Brill and Lindquist~\cite{Brill63})

On the initial time-symmetric surface, the throats are minimal area
surfaces. As the
initial data parameter $\mu$ is varied, the shape of the initial
apparent horizon varies.  If the holes are close enough together
(small $\mu$), a new minimal surface appears, surrounding both black
holes on the initial slice. Cadez~\cite{Cadez74} has calculated the
critical value $\mu_c=1.362$ when this occurs.  We know that event
horizons lie outside of, or are coincident with the outermost trapped
surface~\cite{Hawking73}.  This implies that for values of $\mu$ less
than or equal to $\mu_c$ the initial data is a single
distorted black hole.  As we discuss in
{}~\cite{Anninos93b,Anninos94b,Anninos94f} we have determined the $\mu$
required for a single connected
event horizon by integrating photons through the
spacetimes.  We find that for values of $\mu$ greater than about 1.8
the black holes do not have a common event horizon on the initial
timeslice.

\section{theoretical framework}
\label{sec:theoretical}

We use the 3+1 (or ADM) formalism~\cite{Arnowitt62} to write the
general 4-metric as
\begin{equation}
ds^2=-(\alpha^2-\beta^i\beta_i)dt^2+2\beta_idx^idt+\gamma_{ij}dx^idx^j,
\label{4metric}
\end{equation}
where $\alpha$ is the lapse function foliating the four
dimensional spacetime with three dimensional spatial hypersurfaces
$\gamma_{ij}$, and $\beta^i$ is the shift vector that specifies
three dimensional coordinate transformations from time slice to
time slice.
Throughout this work we use Greek indices (ranging from 0 to 3) to
label four dimensional coordinates, and Latin indices (ranging from 1
to 3) to label spatial coordinates. We use geometric units
in which the gravitational constant and the speed of light are set
to unity.

In the 3+1 formalism,
the vacuum Einstein equations reduce to four constraint equations
\begin{equation}
R-K_{ij}K^{ij}+K^2=0,
\label{ham}
\end{equation}
\begin{equation}
\nabla_j(K^{ij}-\gamma^{ij} K)=0,
\label{mom}
\end{equation}
and twelve evolution equations
\begin{eqnarray}
\partial_t \gamma_{ij}=&&-2\alpha K_{ij}+\nabla_i\beta_j+\nabla_j\beta_i,
\label{evolg}  \\
\partial_t K_{ij}=&&-\nabla_i\nabla_j\alpha +\alpha\left(R_{ij} +K_{ij}K
                -2K_{ik}K^k_j\right) \nonumber \\
                 &&+\beta^k\nabla_k K_{ij}
                +K_{kj}\nabla_i \beta^k +K_{ik}\nabla_j \beta^k.
\label{evolk}
\end{eqnarray}
Here $R$ is the Ricci scalar formed from the spatial 3-metric
$\gamma_{ij}$,
$K_{ij}$ is the extrinsic curvature,
$K$ is the trace of $K_{ij}$
and $\nabla_j$ is the covariant derivative with respect to $\gamma_{ij}$.

\subsection{The Coordinate Systems}
\label{subsec:coordinate}

Because the spacetimes we work with possess an axial Killing vector
(which we set to be $\partial/\partial x^3$),
all variables are independent of $x^3$.
Denoting the azimuthal angle by $x^3\equiv \phi$ and
using the standard ($z$,$\rho$,$\phi$) cylindrical coordinates,
we write the 3-metric for a general axisymmetric
spacetime as
\begin{equation}
\gamma _{ij}=  \Psi^4 \, \hat{\gamma}_{ij}= \Psi^4
               \pmatrix{a  &  c  & 0                \cr
                        c  &  b  & 0                \cr
                        0  &  0  & \rho^2 d \cr }
\label{3metriczr}
\end{equation}
in the coordinate order
$x^i=\left(z,\rho,\phi\right)$.
The variables $a$, $b$, $c$ and
$d$ are functions of the coordinates $z$, $\rho$ and $t$ and are
assumed to be asymptotically flat. The conformal factor $\Psi$ is
a function of $z$ and $\rho$ only and does not evolve in
time.  It is determined on the initial time slice to satisfy the
Hamiltonian constraint (\ref{ham}) and for the type of initial
data that we consider, $\Psi$ and all its derivatives are known
analytically, given by equations (\ref{mPsizr1}) and (\ref{mPsizr2}).
At the initial time slice, the 3-metric (\ref{3metriczr}) takes
the form of the time symmetric Misner data set (\ref{mmetriczr}) with
$a=b=d=1$, $c=0$ and $\Psi=\Psi_M$.

The Einstein equations are simplified when a conformal factor is
introduced into the extrinsic curvature in a manner similiar
to the 3-metric (\ref{3metriczr}).
We define the following form for the
extrinsic curvature
\begin{equation}
K _{ij}=  \Psi^4 \, \hat{K}_{ij}= \Psi^4
               \pmatrix{h_a  &  h_c  & 0                \cr
                        h_c  &  h_b  & 0                \cr
                          0  &    0  & \rho^2 h_d \cr }.
\label{3curvzr}
\end{equation}
The evolution equations (\ref{evolg}) and (\ref{evolk}) can now be
formulated as evolution equations for the metric components
($a$, $b$, $c$, $d$) and their corresponding
curvature components ($h_a$, $h_b$, $h_c$, $h_d$).

As we discussed in section~\ref{sec:initial}, the initial data we
consider consists of two throats connecting two isometric sheets.  By
construction the throats are spheres, centered on the axis along
$\rho=0$,
on which boundary conditions relating the metric across the two
sheets may be imposed.  Since the natural boundaries (the throats and
a sphere surrounding the system far from the throats) do not lie along
constant ($z$,$\rho$) coordinates, it is useful to introduce the
``quasi-spherical'' Cadez \cite{Cadez71} coordinates ($\eta$,$\xi$) with
$\eta$ being a logarithmic ``radial'' coordinate and $\xi$ an ``angular''
coordinate.  Cadez coordinates are related to cylindrical coordinates
through the complex transformation
\begin{eqnarray}
\chi(\zeta)=&& \eta + i \xi \nonumber \\
       =&&\frac{1}{2}\left[\ln(\zeta+\zeta_0)+\ln(\zeta-\zeta_0)\right]
        \label{cadezcoords}   \\
        &&+\sum_{n=1}^{\infty}C_n\left(\frac{1}{(\zeta_0+\zeta)^n}
           +\frac{1}{(\zeta_0-\zeta)^n}\right), \nonumber
\end{eqnarray}
where $\zeta=z+i\rho$, and $\zeta_0=\coth\mu$
is the value of $\zeta$ at the
throat center.  The constant
$\eta$ and $\xi$ coordinate lines of (\ref{cadezcoords}) lie along the
field and equipotential lines of two equally charged metallic
cylinders located at the centers of the two throats $z=\pm
\coth{\mu}$.  The coefficients $C_n$ are determined by a least-squares
method to set the throats (defined by $\rho_{th}^2+(z_{th}\pm
\coth{\mu})^2=1/\sinh^2{\mu}$) to lie on an $\eta=\eta_0=$ constant
coordinate line.  Both $\eta_0$ and the different $C_n$ are computed
on a problem specific basis
for different $\mu$ using this least squares
procedure.  As the $C_n$ are rapidly converging, the series
(\ref{cadezcoords}) can be truncated to low order.  In our simulations
we typically keep only terms up to $n \sim 15$.

The constant Cadez coordinate lines as viewed in
the cylindrical coordinate system
are shown in Fig.~\ref{fig1}a.
The grid
in the Cadez coordinates is shown in the accompanying
Fig.~\ref{fig1}b.
The advantage afforded by this set of coordinates is that they are
spherical near the throats of the black holes and also far away in the
wave zone, thus allowing us to deal with throat boundaries and
asymptotic wave form extractions in a convenient way.
Also, because the natural
boundaries of the spacetime now lie on constant ($\eta$, $\xi$) lines,
our spacetimes are constructed and evolved in a manner
similar to the single distorted black hole spacetimes in references
\cite{Anninos93c,Bernstein93b,Abrahams92a}.
The disadvantage is that
the transformation (\ref{cadezcoords}) introduces a singular saddle
point at the origin ($z=\rho=0$) that is not present in cylindrical
coordinates.  This creates certain numerical difficulties that we
discuss in more detail in section \ref{sec:coordinate}.

The transformation (\ref{cadezcoords}) satisfies the following
Cauchy-Riemann conditions
\begin{eqnarray}
\eta_{,z}=\ \ \xi_{,\rho},&\qquad &\eta_{,zz}
         =-\eta_{,\rho\rho}=\xi_{,z\rho}
          \nonumber \\
\eta_{,\rho}=-\xi_{,z},   &\qquad &\eta_{,z\rho}
            =-\xi_{,zz}=\xi_{,\rho\rho}.
\label{cauchy}
\end{eqnarray}
We note that these first and second derivatives of
$\eta$ and $\xi$ can be computed ``analytically'' (to machine
precision) from
(\ref{cadezcoords}).
The Jacobian of the two coordinate systems after the Cauchy-Riemann
conditions have been applied becomes
\begin{equation}
J=\left(\partial\eta/\partial\rho\right)^2+
  \left(\partial\eta/\partial z\right)^2.
\label{jacobian}
\end{equation}

In this set of coordinates, we define the analog of the
cylindrical based 3-metric
(\ref{3metriczr}) as
\begin{equation}
\gamma _{ij}=  \Psi^4 \, \hat{\gamma}_{ij}= \Psi^4
               \pmatrix{A  &  C  & 0                \cr
                        C  &  B  & 0                \cr
                        0  &  0  & \sin^2\xi ~D \cr },
\label{3metricex}
\end{equation}
with the corresponding extrinsic curvature
\begin{equation}
K _{ij}=  \Psi^4 \, \hat{K}_{ij}= \Psi^4
               \pmatrix{H_A  &  H_C  & 0                \cr
                        H_C  &  H_B  & 0                \cr
                          0  &    0  & \sin^2\xi ~H_D \cr },
\label{3curvex}
\end{equation}
in the order [$\eta$, $\xi$, $\phi$].
The Misner initial data (\ref{mmetriczr}) can be written in the form
of (\ref{3metricex}) using Cadez coordinates as
\begin{equation}
ds^2=\Psi_C^4\left(d\eta^2+d\xi^2+\sin^2\xi \frac{J \rho^2}{\sin^2\xi}
      d\phi^2\right),
\label{mmetricex}
\end{equation}
and identifying $A=B=1$,
$C=0$, $D=J\rho^2/\sin^2\xi$ and
$\Psi=\Psi_C = \Psi_M/J^{1/4}$.

The success of our methodology depends critically on using both sets
of coordinate systems (cylindrical and Cadez) to advantage as we
discuss in section \ref{sec:coordinate}.

\subsection{The Lapse and Shift}
\label{subsec:gauge}

Kinematic conditions for the lapse function $\alpha$ and
shift vector $\beta^i$
complete the
set of Einstein equations
(\ref{ham}) through (\ref{evolk}).
Even though the 3--metric is fixed on the initial slice by the
Hamiltonian constraint, the lapse and
shift can be chosen
arbitrarily on the initial slice and thereafter.

We impose the maximal slicing condition
\begin{equation}
K=\partial_t K=0
\label{maxslice}
\end{equation}
throughout the evolution.
Taking the trace of Eq.~(\ref{evolk}) and
inserting Eq.~(\ref{maxslice})
results in the following elliptic
equation for $\alpha$
\begin{equation}
\nabla^l\nabla_l \alpha =\alpha K_{ij} K^{ij},
\label{lapse}
\end{equation}
where we have used the Hamiltonian constraint to replace
$R$ with $K_{ij}K^{ij}$.
We choose to solve the lapse equation in the nonsingular
cylindrical coordinate system for improved accuracy over Cadez
coordinates.  Solutions to the lapse equation written in
cylindrical coordinates tend to be smoother and better behaved
near the saddle point than solutions obtained by solving the equation
written in Cadez coordinates.  Since the evolution equations for
the extrinsic curvature components contain second derivatives of the
lapse, smoothness in this sensitive region is very important.

For the initial
lapse we have tried both $\alpha=1$,
whereby observers are initially freely
falling, and the solution of Cadez \cite{Cadez71}
\begin{equation}
\alpha\Psi_M=1+\sum_{n=1}^\infty(-1)^n\frac{1}{\sinh{n\mu}}\left(
\frac{1}{{}^{+}r_n}+\frac{1}{{}^{-}r_n}\right),
\label{cadezlapse}
\end{equation}
which is a generalization of the standard static
Schwarschild slicing for a single black hole, but here the system is
not static.
In the first case ($\alpha=1$) the lapse is symmetric across the
throat. In the second case it is antisymmetric and hence equal to zero
on the throat.
We find the antisymmetric
lapse tends to work better as it ``freezes'' the evolution at the throat
and slows it down in regions near and between the throats where
calculations can be troublesome.  For this reason, we work exclusively
with antisymmetric boundary conditions for the lapse across the throat.

The shift vector $\beta^i$
is defined by imposing
\begin{equation}
C = \partial_t C = 0
\end{equation}
to maintain a diagonal 3-metric in the Cadez
coordinates. This condition applied to Eq.~(\ref{evolg}) results in the
differential equation
\begin{equation}
B\frac{\partial\beta^{\xi}}{\partial\eta} +
A\frac{\partial\beta^{\eta}}{\partial\xi} =2\alpha H_C,
\label{shift}
\end{equation}
for the nonvanishing shift vector components $\beta^\eta$ and $\beta^\xi$.
We can rewrite Eq.~(\ref{shift}) by introducing a shift potential
$\Omega$ that satisfies~\cite{Bernstein93b}
\begin{equation}
\beta^\eta=\frac{\partial\Omega}{\partial\xi}, \qquad
\beta^\xi =\frac{\partial\Omega}{\partial\eta}
\end{equation}
to get a single elliptic
equation for $\Omega$
\begin{equation}
B\frac{\partial^2\Omega}{\partial\eta^2} +
A\frac{\partial^2\Omega}{\partial\xi^2} =2\alpha H_C.
\label{shiftpotential}
\end{equation}
We emphasize that it is the off-diagonal Cadez metric component $C$
that is required to vanish and not the off-diagonal cylindrical
component $c$.  In general, the cylindrical metric is nondiagonal.
This choice of gauge for the shift vector proved to be critical to the
overall stability of the numerical evolution, particularly in
suppressing the axis instability.

\section{``patched'' coordinates}
\label{sec:coordinate}

We have investigated a number of different numerical schemes to solve
the problem of colliding two black holes head-on.  The basic idea that
evolved from our progression of trials is to solve for the Cadez
metric and extrinsic curvature components, defined by
Eqs.~(\ref{3metricex}) and (\ref{3curvex}), on the Cadez grid and set
$C=\partial_t C=0$.  This approach has the advantage that the 3-metric
is diagonal which helps to suppress the axis instability and
simplifies the equations of evolution and the extraction of invariant
gravitational waves in the far field.

This approach has the further advantage that it is possible to define
variables for the two black hole system that obey the same evolution
equations with similar boundary conditions, as the single distorted
black hole code developed in previous
work~\cite{Anninos93c,Bernstein93b}.  In fact, the two black hole code
in its final incarnation evolved from the code we developed for
distorted axisymmetric single black hole spacetimes and much of the
discussion in~\cite{Anninos93c,Bernstein93b} is directly applicable to
the two black hole code.  In this section we concentrate on those
techniques developed and modified specifically for the two black hole
problem and in particular the methods we use to overcome difficulties
associated with the singular saddle point present in the Cadez
coordinates.  Further details relevant to both codes may be found in
\cite{Anninos93c,Bernstein93b}.

\subsection{The Grid}
\label{subsec:grid}

Our spacetimes are equatorially plane symmetric, axisymmetric and
isometric through a throat boundary.  The computational grid is
covered with Cadez coordinates and is bounded by the equator ($z=0$),
the axis ($\rho=0$) and the isometry surface ($\eta=\eta_0$).  The
outer boundary is set at $\eta=\eta_{max}$ defined by
$\eta_{max}-\eta_0 =5.8$, which corresponds to an equivalent
Schwarzschild radius ranging from $\sim 125M$ to $\sim 425M$
for $\mu=1.2$ and $\mu=3.25$ respectively.
The outer boundary is
sufficiently far from the two throats
that we can safely impose static conformal flatness boundary
conditions on the metric and extrinsic curvature components in the
asymptotic far field.

Other boundary conditions are specified in an analogous way to the
single black hole spacetimes described in references
\cite{Anninos93c,Bernstein93b}.  Specifically, we require for all the
Cadez grid based variables (except $C$, $H_C$ and $\beta^\xi$) to be
symmetric across the axis and equator. $C$, $H_C$ and $\beta^\xi$ are
antisymmetric across both the axis and equator.  The throat isometry
takes the form
\begin{equation}
\partial_\eta A=
\partial_\eta B=
\partial_\eta D=
\partial_\eta H_C=
\partial_\eta \beta^\xi=0
\end{equation}
and
\begin{equation}
\alpha=C=H_A=H_B=H_D=\beta^\eta=0
\end{equation}
at $\eta=\eta_0$.

These choices result from establishing a consistent convention to
satisfy the Einstein equations subject to the symmetries and
isometries of the problem. The choice of lapse (\ref{cadezlapse}) for
the initial data removes the freedom available to choose an isometry
sign for the metric components and the boundary symmetric constraints on
the shift help preserve the coordinate positions of the axis, equator
and throat boundaries~\cite{Anninos93c,Bernstein93b}. Boundary values
for variables defined in the cylindrical coordinate basis are obtained
from the tensor transformation of the corresponding Cadez grid
variables.

We make one final comment regarding the grid: the grid is comoving
with the black holes throughout the entire evolution.  The dynamical
history is carried solely by the metric components.  As the black holes
approach each other the grid cells do not track the black holes and
squeeze together in the impact area between the two throats.  Instead
the conformal metric function $\hat\gamma_{\eta\eta}=A$, for example,
goes to zero, signifying that the proper distance between grid lines
is decreasing.

\subsection{The Saddle Point}
\label{subsec:saddle}

The most difficult problem associated with the Cadez coordinates
is the coordinate
singularity at the origin ($z=\rho=0$), clearly evident in
Fig.~\ref{fig1}a.
The difficulty with this point, which is inside the
computational domain, is two fold:  ({\it i})~The transformation
Jacobian~(\ref{jacobian}) between the Cadez coordinates
$(\eta$, $\xi$) and
cylindrical coordinates ($z, \rho$) goes to zero at the
saddle point as $(z^2+\rho^2)K_o$, where $K_o$ is a constant
defined in the
neighborhood of the saddle point.  Note that the Jacobian between
cylindrical coordinates and Cartesian coordinates goes to
zero just as $\sim \rho$ at the saddle point, so that Cadez
coordinates are singular with respect to Cartesian coordinates
at $z=\rho=0$.  ({\it ii})~The $\eta$ axis turns abruptly with
respect to the local three geometry at the saddle point.
The $\eta$ axis parallels the $z$ axis
for $\eta <\eta_s$ while becoming the $\rho$ axis for $\eta >\eta_s$,
where $\eta_s$ is the $\eta$ coordinate passing through the origin.

The problem ({\it i})
associated with the vanishing of the Jacobian (volume
element) is a familiar one.  Similar problems exist in
spherical and cylindrical coordinates at the origin.  The problem is
often treated by rescaling variables, factoring out the Jacobian in the
numerical evolution.  The second problem ({\it ii}) is unfamiliar and more
troublesome.  As the spatial three geometry is symmetric to reflection
about the $z=0$ plane, and to rotation about the $\rho$ axis, the time
development of the geometries along the $\rho$ axis and the $z$ axis
are intrinsically different.  The holes are falling towards each other
in the $z$ direction.  Although the initial data is smooth along the
$\eta$ axis, as soon as the holes start falling towards each other, the
discontinuity in the metric functions, say
$\hat\gamma_{\eta\eta}=A$, will
develop at $\eta =\eta_s$. For $\eta <\eta_s$,
$\hat\gamma_{\eta\eta}$
decreases as the proper distance between grid points decreases.
(In fact, there are two competing effects:
the decrease in the proper distance between the
holes, and the grid stretching effects whereby
grid points along the $\eta$
axis with $\eta <\eta_s$ fall towards the ``north'' hole.)  On the
other hand, $\hat\gamma_{\eta\eta}$ increases for $\eta >\eta_s$ as
grid points are falling towards the saddle point, again due to the
grid stretching effect.  The anisotropy in the $\rho$ and $z$
directions at the saddle point translates into the discontinuity along
the $\eta$ axis at $\eta_s$.  The functions
$\hat\gamma_{\eta\eta}$ on the
two sides of $\eta_s$ are really two different geometric objects.
The spatial derivative $\partial_\eta \hat\gamma_{\eta\eta}$ is undefined
analytically at $\eta=\eta_s$ on the $\eta$ axis, and is very large in
the finite differencing approximation.  Likewise the finite
differencing $[\hat\gamma_{\eta\eta}(\eta =\eta_s+\Delta\eta$,
$\xi\approx{\pi\over 2})-\hat\gamma_{\eta\eta}(\eta =\eta_s-\Delta\eta$,
$\xi\approx{\pi\over 2})]/(2 \Delta\eta)$ for grid points off the $\eta$
axis but near the saddle point will also be large.

To see this
problem explicitly, we plot in Fig.~\ref{fig:discontinuous} the
extrinsic curvature function $\hat K_{\eta\eta}=H_A
=-(\partial_t\hat\gamma_{\eta\eta})/\alpha$
evolved for just one time
step.  In this case, since the initial data and all spatial
derivatives are known analytically, the extrinsic curvature can be
computed analytically on the first time step (up to discretization in
time), which can then be used to compute the metric function on the
second time step without computing any numerical spatial derivatives.
In this sense, the metric is known ``analytically'', up to a finite
difference in time.  But the effect of the discontinuity at the origin
($\eta=\eta_s \sim 0.05$, plotted as a circle in
Fig.~\ref{fig:discontinuous})
is clear immediately.
It is easy to see that the same difficulty exists for other quantities
$A, B, H_B...$ {\it etc.}  Without special
precautions or treatments, the evolution quickly contaminates
with numerical noise.
It is possible that certain gauge conditions, such as
quasi-isotropic~\cite{Evans86} or minimal distortion~\cite{Smarr78b},
could minimize the effect of these discontinuities, but we have not
tried them.  The method we use in dealing with this problem is
discussed in the following section.

\subsection{Numerical Issues}
\label{subsec:issues}

The most critical numerical issue is the treatment
of the saddle point and
the region near it.  To help reduce this
problem somewhat,
we construct the Cadez coordinates in such a way as to avoid
placing a grid point
on the origin.  Lines of constant $\xi$ are staggered across the
axis of symmetry and the equator so that both
axis and equator lie on the
half zones.  Lines of constant $\eta$ are set relative to the throat
position $\eta =\eta_0$ at constant discrete spacing $\Delta\eta$
chosen to maintain a grid aspect ratio that is nearly unity.  We do
not enforce a constant $\eta$ line to intersect the saddle point.
Hence, the origin is straddled by both the angular and radial grid
lines (see Fig.~\ref{fig1}a).
Although this procedure eliminates some spurious
effects arising from the singular point, it does not resolve the
problem completely.
Gradients in this vicinity are extremely large, for the reasons
discussed above.

The basic approach we use to evolve data near the saddle point is to
take advantage of the fact that the cylindrical spacetime metric
components (\ref{3metriczr}) are smooth everywhere, including the
saddle point (although the volume element in cylindrical coordinates,
or equivalently the Jacobian to Cartesian coordinates, is still zero at
that point and along the $z$ axis).  We can therefore define a
cylindrical coordinate ``patch'' to cover regions near the saddle point
in the singular Cadez system.  The patch is constructed beginning at
the angular coordinate value $\xi=\pi/2$ and extended along the
angular direction towards $\xi=0$ for a number of zones, depending on the
angular resolution.  The patch is also extended along the radial
direction from the throat all the way out to the outer boundary to
minimize distortions that might be suffered by radially propagating
structures if they encounter patch boundaries ``head-on''.
In this patch region we evolve the cylindrical coordinate based metric
(\ref{3metriczr})
and extrinsic curvature components (\ref{3curvzr})
on the Cadez grid and transform the
solutions via the general tensor relations $T'_{ij}=(\partial
x^k/\partial x'^i)(\partial x^l/\partial x'^j) T_{kl}$ to reconstruct
the Cadez components.

We stress that this scheme is {\it not} a coordinate patch in the
formal sense.  Grid lines are no where laid along the ($z, \rho$)
coordinates, and derivatives are not taken in ($z, \rho$)
coordinates.  Rather, we are simply evolving two sets of components,
Cadez and cylindrical, independently of each other (except for the
coupling at the patch boundaries) on a single Cadez grid. The
nonsingular cylindrical components are used to correct the singular
Cadez components in the patched region.  On the other hand,
the Cadez components
provide corrections to their cylindrical counterparts
everywhere else, helping to suppress the
axis instability that is inherently present in the cylindrical
coordinate system possessing a nondiagonal metric.
To help integrate the patched region into the rest of the spacetime
for smooth evolutions, we construct a layer of buffer zones
surrounding the patched region.  Within this boundary of zones, both sets
of components are evolved and a linear weighting scheme is used to
blend all evolved variables to the values at the edges of this buffer
domain.

In a typical calculation, the lapse collapses approximately
``spherically'' along constant $\eta$ lines, thereby freezing the fields
interior to this domain. (This is the generic behavior of the maximal
slicing condition or any singularity avoiding lapse function.)  Hence,
it is necessary to evolve with the patch in place only until the lapse
drops below a critical value (typically $\sim 0.025$)
at the origin to prevent any
evolution from ocurring there.
Once the lapse falls
below this value, the simulation is continued by evolving only the
Cadez variable components over the entire Cadez grid, including the
saddle area.  The removal of the patch at relatively late times is
necessary to maintain stable and accurate evolutions.  Evolving the
natural Cadez metric components on the Cadez grid does not suffer from
the numerical instabilities inherited from applications of chain rule
derivatives in regions of extreme gradients.

We end this section with a discussion of
an alternative strategy that we tried for the numerical evolution.
Although it was unsuccessful, it is instructive to point
out why it failed.  Given that the cylindrical metric variables are
well behaved near the saddle point, one might consider evolving the
entire system in those variables.  It is well known that the axis
instability can be suppressed by choosing the
shift vector ($\beta^{\rho}$,$\beta^z,0$) such that $c=0$.  This
strategy is effective at minimizing problems related to both the axis
instability and the saddle point, but it introduces a new problem
related to grid stretching.  As noted above, very large gradients
develop in the radial metric function surrounding the hole.  This
radial metric function is composed of the three cylindrical metric
functions $\gamma_{zz}$, $\gamma_{\rho z}$, and $\gamma_{\rho \rho}$
that are actually being evolved.  Using a shift
that forces $c=0$ in this coordinate system causes extreme
angular gradients to develop near the transition between where
$\gamma_{zz}$ is primarily radial (along the $z$-axis) and where
$\gamma_{\rho \rho}$ is primarily radial (along the equator).
Therefore, along a line of about 45 degrees between the axis and
equator, instabilities develop as the grid stretching becomes severe
and the metric functions $\gamma_{zz}$ and $\gamma_{\rho\rho}$
develop stepfunction-like discontinuities.
A nonvanishing $\gamma_{\rho z}$ component serves to absorb
some of this shear but
introduces instabilities on the axis.
For this reason, we adopted
the hybrid scheme described above, whereby the Cadez
metric variables are evolved over much of the grid, and the
cylindrical metric variables are evolved over a smaller
region covering the
saddle.  In this way we were able to benefit from the advantages
afforded by each coordinate system,
while minimizing the problems that each presents.

\section{Numerical Methods and Code Tests}
\label{sec:methods}

\subsection{Solving the Discrete Einstein Equations}
\label{subsec:solving}

The numerical integration of the evolution equations (\ref{evolg}) and
(\ref{evolk}) is performed in an
unconstrained manner because of practical computational time
limitations. We do not enforce either the Hamiltonian (\ref{ham}) or the
momentum constraints (\ref{mom}) during the course of evolution except
at the initial time slice.

Eqs.~(\ref{evolg}) and (\ref{evolk}) are solved using the standard time
explicit second order accurate leap frog method whereby the extrinsic
curvature components are staggered by a half step in time relative to
the metric components.  Schematically we have
\begin{equation}
\gamma_i^{n+1/2}=\gamma_i^{n-1/2}-\left\{2\alpha_i^n K_i^n
                 -\nabla\beta_i^n\right\}\Delta t,
\end{equation}
and
\begin{eqnarray}
K_i^{n+1}=&&K_i^n+\alpha_i^{n+1/2}\left(K_i^{n+1/2}+
          R_i^{n+1/2}\right) \Delta t \nonumber \\
       &&+\left( \beta_i^n\nabla K_i^n +K_i^n\nabla\beta_i^n
         -\nabla\nabla\alpha_i^{n+1/2} \right)\Delta t,
\end{eqnarray}
where subscripts $i$ (superscripts $n$) refer to discrete spatial
(temporal) positions.  To maintain second order accuracy, variables
are extrapolated to the proper time levels as needed using the second
order formula
\begin{equation}
K_i^{n+1/2}=\frac{3}{2} K_i^n -\frac{1}{2} K_i^{n-1}.
\end{equation}
This method propagates gravitational waves with less dissipation and
dispersion than other methods we have tried~\cite{Bernstein93b}.

Explicit methods require stringent restrictions on the size of
timesteps to maintain stability. A condition we found to provide a
good balance between computational speed and accuracy is $\Delta t =
4M\Delta\eta$, where $M$ is half the ADM mass defined
by (\ref{tmass}) and $\Delta \eta$ is the spacing interval
in the radial direction.

As an added measure of stability, we introduce numerical diffusive
terms to the discrete evolution equations.  This effectively adds to
the right-hand-side a term of the form $k\nabla^2 \gamma_{ij}$ and
$k\nabla^2 K_{ij}$ to the evolution equations (\ref{evolg}) and
(\ref{evolk}) respectively.  The coefficient $k$ is chosen as small as
possible to minimize its effect on the accuracy of solutions while
enhancing the stability of the time integration. We construct $k$
dimensionally in the manner $k \sim (\Delta x)^2/(2\Delta t)$ to scale
appropriately with the grid parameters. The proportionality constant
is typically chosen to be of order $\le 0.05$. We have verified that
the addition of these diffusive terms has little effect on the extracted
radiation waveforms for the dominant
$\ell = 2$ modes (but the higher frequency $\ell =
4$ is more sensitive to this procedure, as discussed below).

Spatial derivatives appearing as source terms in the discrete time
evolution equations are differenced using both standard second and
fourth order center differences to approximate $\partial_\eta$ and
$\partial_\xi$ on the uniform Cadez grid.  We find that fourth order
center differences provide more accurate solutions evident in
calculations of apparent horizon masses and gravitational wave form
extractions.
However, because black hole simulations can develop
large gradients, fourth order differences are generally less stable
than using second order centered differences.
(See also Refs.~\cite{Bernstein93b,Bernstein93a} for
discussions of the effect of second and fourth order spatial
derivatives.)
The results presented
in this paper and in the series of companion papers
\cite{Anninos93b,Anninos94b,Anninos93a} were obtained by
center differencing to fourth order all spatial Cadez derivatives
appearing in the evolution Eqs.~(\ref{evolg}) and (\ref{evolk}).
Within the coordinate patch, it is necessary to compute spatial
derivatives with respect to $z$ and $\rho$ of the cylindrical
coordinate based variables.  Finite differences of such derivatives
($\partial_z a$, $\partial_\rho a$,...etc.)  are
computed on the Cadez grid using the chain rule formulas
($\partial_\rho =\eta_{,\rho}\partial_\eta
+\eta_{,z}\partial_\xi$, ...{\it etc}).

The elliptic equations (\ref{lapse}) and (\ref{shiftpotential}) for
the lapse and shift potential respectively are discretized using
second order center differences.  The resulting coupled algebraic
equations give rise to large sparsely banded matrices which we solve
using an iterative two dimensional multigrid algorithm developed by
Steve Schaffer~\cite{Schaffer92}.

\subsection{Convergence Studies}
\label{subsec:convergence}

There are limitations to the grid resolutions that can be achieved.
First, if the angular spacing $\Delta\xi$ is too small, the zones
poised next to the axis can trigger the axis instability during the
evolution.  Second, the phenomenon of ``grid sucking'' in which the
black hole absorbs coordinate lines throughout the evolution is
enhanced with resolution.  As the resolution is increased, the peaks
corresponding to the increased proper distance between nodes on the
grid near the horizon
become more pronounced and gradients grow ever sharper as one
obtains more accurate solutions.  These sharp features are difficult
to resolve numerically and are ultimately responsible for developing
errors at late times, causing the code to crash.  With our fixed mesh
we cannot afford the computer time to add finer zones arbitrarily to
accurately resolve the peaks with sufficient radial resolution to
suppress numerical instabilities for the duration of a simulation.
The convergence studies presented in this article include
100 (27), 200 (35), and 300 (55)
radial (angular) zones on a uniformly spaced grid.

For each parameter run (different values of $\mu$) we extract both the
$\ell = 2$ and $\ell = 4$ waveforms at radii of $30, 40, 50, 60,$ and
$70M$.  (Coordinate positions corresponding to physical distances in
units of $M$ are approximated from the initial data in the
asymptotically spherical far field as
$r\sim \sqrt{\gamma_{\xi\xi}}/M=\Psi^2/M$.)
We use results at each of these radii to
check the propagation of waves and the consistency of our energy
calculations.
Table~\ref{table1} shows the energy radiated across the five
detectors at different grid resolutions for the case $\mu=2.2$.
The convergence rate is at least quadratic in the grid spacing for
all detectors. More specifically,
we find the convergence between the 200 and 300 radial zone evolutions
to be in the range 0.2--0.7\% for each individual detector, with
the larger deviations occurring for the innermost and outermost
detectors. The median deviation is $\sim$ 0.4\%.
We also find
a general agreement among the different detectors at the same resolution.
For example, the $rms$ deviation (taken over all detectors)
from the average radiated energy
is $\sim$ 5\% for the 200 radial zone evolution. A maximum deviation
of $\sim$ 15\%
occurs between the two detectors furthest from one another.
Reasons for the larger discrepancies among the inner- and outer-
most detectors are the following:
({\it i}) the detectors closer in have a
greater difficulty in separating gravitational wave signals from other
parts of the gravitational field, ({\it ii})
the detectors further out are
more susceptible to numerical resolution problems and artificially
induced diffusion, and ({\it iii})
the finite numerical run times allow for
more of the gravitational wave train to pass through the inner
detectors.

In Fig.~\ref{figl2mu2.2} we show the $\ell=2$ waveform extracted at
$r=40M$ for the three different spatial resolutions of 100 (27), 200
(35), and 300 (55) radial (angular) zones
for the case $\mu=2.2$.  Up until $t\sim 150M$ the results are quite
similar in all three cases.  After that time the low resolution run
develops some difficulty due to the poorly resolved peak in the metric
functions and becomes unstable.
We note that with the higher resolution simulations one obtains a
slightly better fit to the known quasinormal waveform, especially at
late times.  The higher resolved
waveforms suffer less dispersion and damping
attributable to numerical effects.
Prior to $t\sim 125 M$,
the waveforms agree to within $\sim$3\% for all resolutions.
Furthermore, the diffusion and patch
parameters are different for each run, so the $\ell=2$ waveforms are
quite stable with respect changes in computational parameters.
In fact for a
fixed resolution, varying the patch parameters
(such as the size of the patch, the time at which it is
lifted and the numerical diffusion) the waveforms vary by no more
than $\sim$10\%, and typically by less than a few per cent.

Next we show results from the more difficult $\ell = 4$ extraction in
Fig.~\ref{figl4mu2.2} for the same $\mu=2.2$ case.
The $\ell=4$ mode is
clearly more sensitive to the computational parameters than
the $\ell=2$ extraction in both
the signal preceding the strong quasinormal
ringing (beginning at $t\sim 75M$) and in
the amplitude of the ringing signal.
Also for a fixed resolution,
the amplitude of the $\ell=4$ waveforms varies by about a factor of
two with large changes in the computational parameters,
with the largest effect coming from the patch parameters.  As
the energy output varies quadratically with the wave amplitude, the
energy carried by the $\ell=4$ modes is uncertain to about an order of
magnitude.  In spite of these effects, a fit of the two lowest
$\ell=4$ quasinormal modes to the high resolution run, shown in
Refs.~\cite{Anninos93b,Anninos94b}, is quite good.

The reasons for the $\ell = 4$ extraction to be less accurate are
clear. In the first place, the amplitude of the $\ell = 4$ component
is much smaller than that of $\ell = 2$, hence harder to extract
from the background noise level.
Moreover the more complicated angular distribution of the $\ell = 4$
component needs more angular zones to be fully resolved than have been
used in these runs.
Even though we are
unable to determine with great certainty the absolute amplitude of the
$\ell = 4$ signal with our present code, it is clearly seen in the
data and does match the expected quasinormal frequency. On the other
hand, we are confident in the accuracy of the larger and more robust
$\ell=2$ signals, which are not sensitive to computational details as
shown in Fig.~\ref{figl2mu2.2}. These figures show both the strengths
and limitations of the present code.

\section{summary}
\label{summary}

In this paper we have presented the methods we developed to evolve
the head-on collision of two equal mass
black holes initially at rest.  This problem
is a difficult one that has required
a number of numerical strategies
designed to handle large gradients, to suppress instabilities, and
to deal with singular points in the coordinate system.
The result
of our work is a code that can accurately evolve the black hole
collision problem for a range of initial data sets.
By considering simulations run with different numerical parameters and
at different resolutions and we have demonstrated that the
$\ell=2$ waveforms and hence the total radiated energy calculations
are accurate. The amplitude and early time behavior of the $\ell=4$
waveforms are more sensitive to numerics, although the essential
features (the wavelength and damping time) of the quasinormal ringing
are clearly present and can be accurately resolved.  In a series of
companion papers~\cite{Anninos93b,Anninos94b,Anninos93a}
we present detailed analyses of the physical results
obtained using the methods outlined in this article.

Our work represents a step towards solving the more general and
dynamic problem of coalescing binary black holes.  It is our long term
goal to develop a multi-purpose three dimensional code capable of
simulating dynamic fully general relativistic spacetimes containing
multiple black holes of arbitrary mass, rotation and impact
parameters.

\acknowledgements
We would like to thank David Bernstein for a number of
helpful discussions.  This work was supported by NCSA, NSF
Grant~91-16682, and NSERC Grant No. OGP-121857, and calculations
were performed at NCSA and the Pittsburgh Supercomputing Center.



\begin{figure}
\caption{(a)  The Cadez grid is constructed for the case $\mu=2.2$
and displayed in a single quadrant with cylindrical coordinates.
The throats are centered on the
axis at $z=\pm \coth\mu$.
Lines of constant $\eta$ concentrically surround the throat locally,
and become spherical far from the holes. (b) The
computational grid is shown in Cadez
coordinates.}
\label{fig1}
\end{figure}

\begin{figure}
\caption{The extrinsic curvature function
$\hat K_{\eta \eta}=H_A$ is plotted along a $\xi=$ constant
line ($\sim \pi/2$)
after a single time step in evolution.  The geometric
discontinuity in this function is evident.}
\label{fig:discontinuous}
\end{figure}

\begin{figure}
\caption{
We show the $\ell=2$ waveform at various resolutions  of 100 (27), 200
(35), and 300 (55) radial (angular) zones
for the case $\mu=2.2$.  Except for the low resolution at very late times,
the waveforms agree to within less than 3\% across all simulations.}
\label{figl2mu2.2}
\end{figure}

\begin{figure}
\caption{
We show the $\ell=4$ waveform at various resolutions of 100 (27), 200
(35), and 300 (55) radial (angular) zones
for the case $\mu=2.2$.  The amplitude of the ringing modes
for these simulations varies by
about a factor of 2, depending upon the grid
resolution and other computational
parameters such as the size and duration
of the numerical patch and the added numerical
diffusion.  In spite of the uncertainty in the amplitude of the
signals, the $\ell=4$ quasinormal mode is clearly
present in all cases.}
\label{figl4mu2.2}
\end{figure}

\begin{table}
\begin{tabular}{|c|c|c|c|} \hline
detector   & 100 radial zones & 200 radial zones & 300 radial zones
\\ \hline
$30M$ & 7.032 $\times 10^{-4}$
      & 6.098 $\times 10^{-4}$
      & 6.068 $\times 10^{-4}$
  \\ \hline
$40M$ & 6.052 $\times 10^{-4}$
      & 5.785 $\times 10^{-4}$
      & 5.773 $\times 10^{-4}$
  \\ \hline
$50M$ & 5.710 $\times 10^{-4}$
      & 5.619 $\times 10^{-4}$
      & 5.606 $\times 10^{-4}$
  \\ \hline
$60M$ & 5.346 $\times 10^{-4}$
      & 5.476 $\times 10^{-4}$
      & 5.461 $\times 10^{-4}$
  \\ \hline
$70M$ & 5.069 $\times 10^{-4}$
      & 5.274 $\times 10^{-4}$
      & 5.313 $\times 10^{-4}$
  \\ \hline
\end{tabular}
\caption{
Convergence study of the total radiated energy for the case
$\mu=2.2$. The energies are computed at the five
detector locations
for three different spatial resolutions. The convergence rate
is at least quadratic in the grid spacing for all detectors
and deviations between the 200 and 300 radial zone simulations
are on the order of a few percent.
\label{table1}}
\end{table}


\begin{thebibliography}{10}

\bibitem{Thorne94a}
K. Thorne,  in {\em Proceedings of the Eight Nishinomiya-Yukawa Symposium on
  Relativistic Cosmology}, edited by M. Sasaki (Universal Academy Press, Japan,
  1994).

\bibitem{Anninos93b}
P. Anninos, D. Hobill, E. Seidel, L. Smarr, and W.-M. Suen, Phys. Rev. Lett.
  {\bf 71},  2851  (1993).

\bibitem{Anninos94b}
P. Anninos, D. Hobill, E. Seidel, L. Smarr, and W.-M. Suen, Physical Review D
  (1994), in preparation.

\bibitem{Anninos93a}
P. Anninos, D. Bernstein, S. Brandt, D. Hobill, E. Seidel, and L. Smarr,
  Physical Review D  (1994), in press.

\bibitem{Cadez71}
A. \v{C}ade\v{z}, Ph.D. thesis, University of North Carolina at Chapel Hill,
  1971.

\bibitem{Smarr75}
L. Smarr, Ph.D. thesis, University of Texas, Austin, 1975.

\bibitem{Eppley75}
K. Eppley, Ph.D. thesis, Princeton University, 1975.

\bibitem{Smarr79}
L. Smarr,  in {\em Sources of Gravitational Radiation}, edited by L. Smarr
  (Cambridge University Press, Cambridge, 1979), p.\ 245.

\bibitem{Smarr76}
L. Smarr, A. \v{C}ade\v{z}, B. DeWitt, and K. Eppley, Physical Review D {\bf
  14},  2443  (1976).

\bibitem{Anninos93c}
P. Anninos, D. Bernstein, D. Hobill, E. Seidel, L. Smarr, and J. Towns,  in
  {\em Computational Astrophysics: Gas Dynamics and Particle Methods}, edited
  by W. Benz, J. Barnes, E. Muller, and M. Norman (Springer-Verlag, New York,
  1994), to appear.

\bibitem{Bernstein93b}
D. Bernstein, D. Hobill, E. Seidel, L. Smarr, and J. Towns, Physical Review D
  (1994), in press.

\bibitem{Abrahams92a}
A. Abrahams, D. Bernstein, D. Hobill, E. Seidel, and L. Smarr, Physical Review
  D {\bf 45},  3544  (1992).

\bibitem{Misner60}
C. Misner, Phys. Rev. {\bf 118},  1110  (1960).

\bibitem{Einstein35}
A. Einstein and N. Rosen, Phys. Rev. {\bf 48},  73  (1935).

\bibitem{Misner63}
C.~W. Misner, Ann. Phys. {\bf 24},  102  (1963).

\bibitem{Lindquist63}
R.~W. Lindquist, Jour. Math. Phys. {\bf 4},  938  (1963).

\bibitem{Brill63}
D.~S. Brill and R.~W. Lindquist, Phys. Rev. {\bf 131},  471  (1963).

\bibitem{Hahn64}
S.~G. Hahn and R.~W. Lindquist, Ann. Phys. {\bf 29},  304  (1964).

\bibitem{Cadez74}
A. \v{C}ade\v{z}, Annals of Physics {\bf 83},  449  (1974).

\bibitem{Hawking73}
S.~W. Hawking,  in {\em Black Holes}, edited by C. DeWitt and B.~S. DeWitt
  (Gordon and Breach, New York, 1973).

\bibitem{Anninos94f}
P. Anninos, D. Bernstein, S. Brandt, J. Libson, J. Mass\'o, E. Seidel, L.
  Smarr, W.-M. Suen, and P. Walker, Physical Review Letters  (1994), submitted.

\bibitem{Arnowitt62}
R. Arnowitt, S. Deser, and C.~W. Misner,  in {\em Gravitation: An Introduction
  to Current Research}, edited by L. Witten (John Wiley, New York, 1962).

\bibitem{Evans86}
C. Evans,  in {\em Dynamical Spacetimes and Numerical Relativity}, edited by J.
  Centrella (Cambridge University Press, Cambridge, 1986).

\bibitem{Smarr78b}
L. Smarr and J. York, Physical Review D {\bf 17},  2529  (1978).

\bibitem{Bernstein93a}
D. Bernstein, Ph.D. thesis, Dept. Of Physics, University of Illinois
  Urbana-Champaign, 1993.

\bibitem{Schaffer92}
S. Schaffer, private communication.

\end{thebibliography}
\end{document}